# Glaciological Window into the Pace of Earth's Organic Carbon Cycle

Matthieu E. Galvez[a]

[a] Institute of Geochemistry and Petrology, Swiss Federal Institute of Technology, CH-8092 Zurich, Switzerland
[b] Geological Institute, Swiss Federal Institute of Technology, CH-8092 Zurich, Switzerland

## Paper metrics

Words: 2990
Figures: 3
Tables: 1

## Significance Statement

Air bubbles trapped in Antarctic ice show that atmospheric oxygen has declined slowly over the last 800.000 years. What this observation means for the relative strength of organic carbon sources and sinks is unknown. Here, using a compilation of C, Fe, S and H flux between the exosphere, continents, and mantle of the Earth, we propose an interpretation of the glaciological record. We demonstrate that the organic carbon cycle is a net source of $O_2$ in the Pleistocene —and equipotent $CO_2$ sink —despite coexisting atmospheric deoxygenation. Our work highlights that remarkably efficient feedbacks tie atmospheric, ocean and lithospheric flux of carbon, iron and sulfur, and provide insights into the long-term evolution of organic carbon reservoirs on Earth.

Abstract

The glaciological record of atmospheric composition suggests $O_2$ has declined over the last 800,000 years at an average rate of $0.3 \times 10^{12}$ mol (Tmol) $O_2$ yr$^{-1}$ [ref (1)]. Because the geological carbon cycle regulates long-term atmospheric oxygen concentrations, fluctuations in atmospheric $O_2$ are typically attributed to an imbalance between the weathering of organic carbon (OC) and reduced sulfur on land — a sink of atmospheric $O_2$ — and the burial of OC and reduced sulfur in marine sediments —a source of $O_2$ (refs. (1, 2)). Here we compile and confront a database of C, Fe, S and H exchanges between the fluid Earth (atmosphere, ocean, biosphere) and lithosphere (crust and upper mantle) with the record of deoxygenation to quantify organic carbon sources and sinks in the Pleistocene. We show that the subduction of oxidized oceanic lithosphere and degassing of reduced gas from the Earth's interior is a sink of oxygen, and that this sink significantly exceeds the rate of atmospheric deoxygenation. This discrepancy requires that the organic carbon cycle was a net source of $O_2$ and sink of $CO_2$ —photosynthesis outpaced respiration by an average of ~40 MtC yr$^{-1}$ over the Pleistocene despite declining $pO_2$. The cost for the relatively invariant atmospheric oxygen concentration is the co-existence of two photosynthetically-driven imbalances in the cycles of iron and carbon that offset each other to near perfection. The weak escape of OC from continents and oceans to the lithosphere is more intriguing. It demonstrates that the organic carbon cycle remains under surprisingly strong kinetic control, despite evolution/optimization of respiratory metabolisms(3) and rising atmospheric $O_2$ for more than 2.4 billion years.

Earth is the only planet that maintains atmospheric $O_2$ concentrations so far from thermodynamic equilibrium (4). Although the mechanism by which molecular $O_2$ accumulated to its present levels remains unclear in detail, this progressive accumulation broadly reflects the coevolution of ecosystems with their environment over geological time (5, 6). But the ratio of $O_2$ to $N_2$ ($\delta O_2/N_2$) in gas bubbles trapped in Antarctic ice indicates that over the Pleistocene, more or less the last 800,000 years, atmospheric $O_2$ has declined at an average rate of 0.3 Tmol yr$^{-1}$ [Fig. 1, ref (1)]. This decline offers a unique opportunity to investigate the meaning and controls on atmospheric $O_2$ fluctuations today, and in the distant past.

Enhanced dissolution of atmospheric oxygen in the ocean during late Neogene cooling is a possible cause for the decline. In accordance with early models of ocean-atmosphere dynamics (7-9), this mechanism could involve a boost in aerobic respiration of labile organic materials dissolved in the deep-ocean ($OC_{ocean}$) and serve as a net sink of $O_2$. Alternatively, if the weathering of reduced carbon ($OC_{petrogenic}$), sulfur and/or iron on continents outpaces the burial flux of reduced materials to marine sediments(2)—a source of $O_2$—, could also result in transient decline of $pO_2$. Indeed, the oxidation of sulfur has seen a recent surge of attention as a potential "hidden" regulator of the cycle of alkalinity, climate (10, 11), and of the continental redox balance(1, 10).

These alternatives illustrate a more fundamental complexity with the oxygen cycle. Carbon, hydrogen, iron and sulfur species of various valence states are linked through redox reactions affecting the crust, ocean, biosphere, and atmosphere simultaneously, and at different rates, all regulating $O_2$ concentration. For this reason, the meaning of $O_2$ trends for the organic carbon cycle and its feedbacks remain equivocal.

Here, we take another approach. We do not focus on the cause of the decline itself; we limit ourselves to examining its implication for the global redox balance. In recent years, extension of the glaciological record of $O_2$ beyond the noise of glacial-interglacial variability —back to about 1 Myr ago (1)—offers a unique opportunity to link surface and deep $O_2$ cycles (12), and shed light on the structure and pace of the organic carbon cycle. Here, we quantify redox exchanges (ie. electron and $O_2$ equivalent fluxes), between surface of the Earth, continents and mantle. This inventory is used to assess the existence, sign, and magnitude of an imbalance in the pre-anthropogenic organic carbon cycle. The methodological innovation of our work is in linking the glaciological and geological records of $O_2$ flux between surface and deep Earth. We demonstrate that OC production pathways (ultimately photosynthesis) has outpaced the two main OC oxidation pathways on Earth in the Pleistocene: the respiration of biological OC ($OC_{bio}$) and the oxidation of petrogenic kerogens ($OC_{petro}$). This means that the carbon cycle has functioned as a net source of $O_2$—and equipotent sink of $CO_2$—despite evidence for atmospheric deoxygenation. More importantly, the similarity between flux of $OC_{petro}$ from continents to oceans and the net $O_2$ export by organic carbon burial suggests that the persistence of an oxygenated atmosphere in the Pleistocene may not be an issue of *supply* (ie. of reductant and oxidant), but rather one of sluggish global OC oxidation kinetics in the environment. Partial graphitization of organic materials may be the most important yet overlooked set of reactions for the long-term kinetics of Earth's organic carbon cycle.

1. Redox neutrality in the geobiosphere

At any given time, variations of atmospheric oxygen are controlled by the overall balance of geological and biological sources and sinks of electrons, converted here to equivalent $O_2$ fluxes [Table 1, Supplementary Information and ref (13)]. The most important sources of electrons (ie. sinks of $O_2$) to the atmosphere-ocean system in the Pleistocene include the heterotrophic respiration of organic matter (14); the weathering of petrogenic carbon in continental rocks (15); the weathering of reduced iron *(*i.e., sulfides and Fe-carbonates) and sulfur (i.e.. sulfides) in continental rocks (16); the hydration/oxidation of Fe-bearing minerals in oceanic basalts and the lithospheric mantle by oxygenated seawater (17) (e.g. serpentinization), and the oxidation of thermogenic methane (18) and reduced volcanic gases (19). The most important sinks of electrons (ie. sources of $O_2$) are the net photosynthetic production of biospheric OC (14) [$OC_{biosphere}$ includes living biomass, particulate (POC) and dissolved organic matter (DOC)], and the burial of reduced kerogen [see Supplementary Information and ref (20)], sulfur (2), and iron (16) in sea floor sediments and in the oceanic lithosphere.

At steady-state, $O_2$ sources and sinks associated to the carbon, iron, and sulfur cycles obey the mass (redox) conservation relation:

$$\Delta_{OC} + \Delta_S + \Delta_{Fe} = \delta^{ao} + \delta^h \tag{1}$$

where $\Delta_{OC}$, $\Delta_S$, $\Delta_{Fe}$ designate net $O_2$ budget linked to variations in size of the organic carbon (biological and geological), reduced sulfur, and reduced iron reservoirs respectively, $\delta^{ao} = \frac{dO_2}{dt}$ is the variation of $O_2$ in the atmosphere-ocean system, and $\delta^h$ is the net $O_2$ budget corresponding to hydrogen lost to space. $\Delta_i = F^{in} - F^{out}$ and thus represent differences between input and output of C, Fe and S. Because $\delta^h$ has been negligible over the Phanerozoic [table 1, ref(21)], it will not be considered further. If one distinguishes the exchange between atmosphere and mantle ($\Delta^s$), atmosphere and continents/sediments ($\Delta^c$) and atmosphere and biosphere-ocean ($\Delta^b$) (Figure 2), equation 2 takes the alternative form:

$$\Delta^s + \Delta^c + \Delta^b = \delta^{ao} \tag{2}$$

Equalities (2) and (3) state that Fe, C, H, and S flows can only oxidize the planet as a whole via hydrogen loss to space (21). If one assumes global redox neutrality ($\delta^h \sim 0$, Eq. 3), however, then long-term variations of atmospheric oxygen ($\delta^{ao}$) reduce to a question of internal $O_2$ partitioning, dependent upon C, Fe, H, and S exchanges between biosphere, ocean, atmosphere and lithosphere. Constraints on those exchange, irrespective of their magnitude, can be derived when $\delta^{ao}$ is known. We exploit this fundamental property to quantify organic carbon sources and sinks ($\Delta_{OC}$) in the Pleistocene.

### 2. Net redox output to the fluid Earth by the organic C cycle

Quantifying organic carbon sources and sinks with Eq. 3 requires prior constraints on $\Delta^s$. The latter is calculated as the difference between input and output of electrons from the mantle to the fluid Earth (atmosphere, ocean, and biosphere) via exchange of reduced C ($C^0$), Fe ($Fe^{2+}$), S ($S^{2-}$, $S^-$, $S^{4+}$), and H ($H^0$) species (22) at subduction zones, mid-ocean ridges, and oceanic islands (for conversion of element flux to electron and/or $O_2$ equivalents, cf Table 1). Uncertainties in element flux between reservoirs, and uncertainties in the concentrations of various valence forms of each elements are propagated via Monte Carlo Simulation. The uncertainties reported in the main text correspond to 1σ. When a skewed distribution is more appropriate (e.g. petrogenic C export, Table 1, Fig 2B), we refer to the maximum value of the distribution (3 σ$_+$) and the minimum value of the distribution (3 σ$_-$).

The main export flux is that of ferrous iron ($Fe^{2+}$, olivine) in the altered oceanic crust and upper mantle (serpentinite) (release of oxidizing power to the AO system of $F^{s,i}_{Fe}$ ~ + 11.4 (±2.5, σ) Tmol $O_2$ yr$^{-1}$; Table 1). The second important export flux is that of reduced sulfur ($S^{2-}$ and $S^-$) in the altered oceanic crust and upper mantle, which contributes $F^{s,i}_S$ ~ + 2.5 (±0.6, σ) Tmol $O_2$ yr$^{-1}$. There is evidence for refractory DOC in the altered oceanic lithosphere, but it may not exceed 0.1 Tmol $O_2$ yr$^{-1}$ (ref. (23)) [Table 1]; it is ignored here. Export fluxes are moderated by inputs of S- and H-bearing rocks, gas and fluids (~ - 3.5 (± 0.6 σ) Tmol $O_2$ yr$^{-1}$, see method), and Fe- bearing rocks in arc and intraoceanic settings (Peridotite, MORB and OIB, $F^{s,o}_{Fe}$ ~ - 14(± 3.0, σ) Tmol $O_2$ yr$^{-1}$). The flux of oxidized iron to the altered oceanic crust (AOC) is computed using compilations of $Fe^{3+}/Fe_{tot}$ ratio for unaltered MORB of 0.16 (± 0.09 σ) % (σ) by Cottrell et al. (24) and Birner et al (25), and of 0.22 (± 0.09 σ) % for AOC (basalt and gabbro). The $Fe^{3+}/Fe_{tot}$ ratio is null for unaltered peridotites, and equals 0.54 (± 0.2 σ) % (σ) for serpentinites (19). We assume that the average thickness of fully serpentinized lithospheric mantle entering trenches worldwide ranges between 0.5 (5% serpentinization over 10km) and 3.4 km (17% serpentinization over 20km) (19), and the structure of the AOC is from(26). Our method requires that inputs and outputs of Fe and S to and from the atmospheric-ocean system balance each other during the hydrothermal alteration process: they are equal to 63.7 ± 11 TmolFe yr$^{-1}$, and 2.2 (± 0.3 σ) TmolS yr$^{-1}$, respectively (cf Supplementary Material). Overall, the budget between input to and output of reductants from the fluid Earth echoes previous inventories(17, 19); the altered oceanic lithosphere is a net source of electron, ie. a net $O_2$ sink of $\Delta^s$ ~ - 3.8 (± 1.9, σ) Tmol $O_2$ yr$^{-1}$ (Table 1). Fe oxidation (~ -2.6 Tmol$O_2$ yr$^{-1}$) contributes for ~65% of this total, whereas the remaining ~35% is due to S oxidation (~ -1.2 Tmol$O_2$ yr$^{-1}$).

This analysis provides constraints on sedimentary cycles of Fe, S and C as a whole via Eq. 2, and this is the most important result of our study: sedimentary cycles have been, collectively and in average, a net source of $O_2$ for the fluid envelopes over the last million year, despite atmospheric deoxygenation. Substituting $\Delta^s$ in Eq.3 gives

$$\Delta^c + \Delta^b = 3.5 (\pm 1.9, \sigma) \text{ Tmol } O_2 \text{ yr}^{-1} \quad (3)$$

where $\Delta^b$ is the average change in size of the biomass over the last million year. Importantly, this result exploits not the absolute value of $\delta^{ao}$, but the fact that it $\delta^{ao}$ is negligible in front of magnitude of rock oxidation pathways at the surface of the Earth.

## 3. Implications for the balance of sedimentary S and C cycles

Discriminating the role of S and C sedimentary (weathering and burial) cycles from $\Delta^c$ is complex because the uncertainties on continental weathering and burial fluxes of S and Fe are large(27). It has been generally assumed that S and Fe should compensate and ($\Delta^c_{S, Fe}$) be close to redox neutrality(16). This seems to be confirmed by recent studies. For example, isotopic proxies(10) indicate that about 42%, and not 20% (28), of riverine sulfate exported to the ocean is due to pyrite oxidation (Fig 2); and Tostevin et al.(29) noted that about 20 to 35 % of the total S input to the ocean precipitates as sedimentary sulfides(29). Therefore, these work imply a burial flux of about 0.6–1.2 Tmol S yr$^{-1}$ if an initial riverine sulfate export of 2 to 3.5 Tmol yr$^{-1}$ is considered; but the latter value too has been debated(30). Regardless, given the relative uncertainties in this estimates, we deem necessary to attribute to this component of the S cycle a large error (Fig. 2). Using Fe weathering fluxes from (27) and an average molar ratio S/Fe = 2 (pyrite) in shelf and deep-sea sediments from (27) gives $\Delta^c_{S, Fe}$ = -0.4 ($\pm$ 1.4, $\sigma$) Tmol $O_2$ yr$^{-1}$ (Table 1). At first sight, this is broadly consistent with (16), but the uncertainty is large. Clearly, more work is needed to improve the consistency between inventories and isotopic estimates of the S cycle.

Regardless, if the decrease of $\delta^{ao}$ is as close to 0 as ref (1) suggests, we suggest that there is a high likelihood that the C cycle remains, as a whole, a net source of $O_2$ (Fig 2). Note that this conclusion does not depend on the absolute C burial flux, or on any of the OC oxidation pathways. It simply derives from the igneous cycles, their uncertainties, and on $\Delta^c_{S, Fe}$ through application of Eq. (2). Therefore, the continental weathering and burial of OC in marine sediments exports $O_2$ to the atmosphere-ocean system—at a rate of 3.8 (± 2.3, σ) Tmol $O_2$ $yr^{-1}$ in the Pleistocene—despite overall deoxygenation.

4.  Imbalance between petrogenic $C_{org}$ oxidation and $C_{org}$ burial

Under some assumption, this result may inform the efficiency of $OC_{petro}$ weathering. For example, the net output of the carbon cycle ($\Delta^c_{OC} + \Delta^b$) obtained previously fixes the difference between the global production ($F^{c,i}_{OC}$) and oxidation ($F^{c,o}_{OC}$) of crustal OC(31) if $\Delta^b$ is known. For sake of numerical illustration, let's assume that the size of the biosphere is constant ($\Delta^b = 0$), although we recognize that this is likely to be incorrect, given large-scale changes in ice-sheet extent and ocean circulation re-organization during the Pleistocene(32). We use the relation $\Delta^c_{OC} = F^{c,i}_{OC} + F^{c,o}_{OC}$. $F^{c,i}_{OC}$ (Table 1) is obtained by subtracting the reburied $OC_{petro}$ fraction $F_{petro}$ = 3.6 (3σ- = 2.1; 3σ+ = 5.1) TmolC $yr^{-1}$ [from ref (33)] from the total marine OC burial flux (ie. ~170(± 20, σ) Mt C $yr^{-1}$ after ref (34, 35), with a correction for diagenetic decarboxylation via a (molar) burial quotient $O_2$:$OC_{bio}$ = 1.1 (Supplementary Material). We obtain $F^{c,i}_{OC}$ = 11.6 ± 3.3 Tmol $O_2$ $yr^{-1}$ which, combined with $\Delta^c_{OC}$, gives a total OC oxidation flux of $F^{c,o}_{OC}$ = -7.7 (±3.5, σ) Tmol $O_2$ $yr^{-1}$. $F^{c,o}_{OC}$ is split between thermogenic $CH_4$ oxidation ($F^{c,o}_{ch4}$) and $OC_{petro}$ oxidation ($F^{c,o}_{petro}$). If $F^{c,o}_{ch4}$ = -3 (±1, σ) Tmol $O_2$ $yr^{-1}$ (Table 1, ref (18)] to cover the broad range of independent estimates [cf. ref(18)], we obtain by difference $F^{c,o}_{petro}$ = -4.8 (±3.6, σ) Tmol $O_2$ $yr^{-1}$ (Table 1). This other consequence of Eq. (2) and (3) is important because it suggests that the efficiency of $OC_{petro}$ oxidation ($\Gamma_{OCpetro} = F^{c,o}_{petro}/(F^{c,o}_{petro} + F_{petro})$) is limited, with a mode at 64%. Yet, this value is obtained by assuming $\Delta^b = 0$; and we therefore concur with (36) that potential long-term fluctuations of the dissolved, terrestrial and biospheric $OC_{bio}$ pool over long timescales may have profound geological implications(36).

5.  Geological implications

The main insight of our work comes from confrontation between geological records of electron flow in the geobiosphere and their expression in atmospheric $O_2$ fluctuations. It provides quantitative constraints on the pace of the carbon cycle in the Pleistocene, and indications on its sensitivity. Especially striking here is that global $OC_{bio}$ production pathways (ultimately photosynthesis) outpaced all OC oxidation pathways (chiefly respiration of $OC_{bio}$, oxidation of

CH$_4$ and OC$_{petro}$ weathering) by an average of ~40 MtC yr$^{-1}$ over the last million years (ie. Δ$_c$ > 0). Erosion-driven oxidation of petrogenic carbon exposed to weathering is overall inefficient with Γ$_{OCpetro}$ ~ 64%. The Fe cycle is imbalanced, ie. Δ$_{Fe}$ < 0, and it redox balances the organic C cycle (Δ$_{Fe}$ + Δ$_{OC}$ ≈ 0). Several important implications follow:

That Δ$_{OC}$ > 0 means that the organic carbon cycle has functioned as a net source of O$_2$ —and sink of CO$_2$—despite atmospheric deoxygenation (Fig. 3). We advance no argument to support or invalidate the deoxygenation *rate* proposed by ref (1). However, we claim that the *only statistically significant* implication of the data available *so far* is that the C cycle has been a sizable net source of O$_2$, *despite* uncertainties in the S cycle. The latter is, within large margins of errors, in near redox balance (Table 1). A net export of OC$_{bio}$ from the atmosphere-ocean-biosphere reservoir to the lithosphere does not, however, imply continuous accumulation of OC in the solid Earth. Mantle upwelling at mid-ocean ridges consumes elemental carbon, as well as mantle oxygen, and produces CO$_2$-bearing basalts and a residuum of reduced abyssal spinel peridotites (ie. decompression/redox melting, ca. 42 MtC yr$^{-1}$). The process results in a mantle locally more reduced (37) according to the reaction:

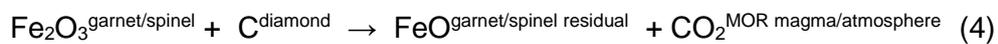

$$Fe_2O_3^{garnet/spinel} + C^{diamond} \rightarrow FeO^{garnet/spinel\ residual} + CO_2^{MOR\ magma/atmosphere} \quad (4)$$

Redox mechanisms analogous to Eq. 4 may similarly lead to partial assimilation of continental OC$_{petro}$ [e.g. graphite (38)] by oxidized arc-magma that rise and reside in the continental lithosphere, returning continental OC to the surface as CO$_2$ (Fig 1, Fig 4).

We also find that Δ$_{Fe}$ + Δ$_{OC}$ ≈ 0, ie. a process in which the oxygen released to the surface by the carbon cycle is subducted/accreted back to its ultimate source (Fig. 2b, Supplementary figure 1)— the lithosphere. Because Fe weathering on the seafloor (Δ$^S_{Fe}$) may depend on oceanic and nutrient dynamics(39), we propose that there may be a link between long-term climatic variability and magnitude of lithospheric weathering, ie. Δ$^s$. The cadence of atmospheric oxygenation is intrinsically tied to the longevity and vigor of O$_2$ deoxygenation pathways; which means that O$_2$ accumulation may have never been irreversible, or only in appearance.

We find that the oxidation efficiency of $OC_{petro}$ is approximately 64%. Fundamentally, that the oxidative pathways of the carbon cycle remain relatively inefficient is surprising given evolution/optimization of respiratory metabolisms (3) and rising atmospheric $O_2$ for more than 2.4 billion years. In practice, our estimate of $\Gamma_{OCpetro}$ matches independent estimates of $\Gamma_{OCpetro}$ at the scale of the Taiwan orogen [~67%, ref (40)]. This match is also surprising because it means that the tropical sub-system mimics the macroscale dynamics of the organic carbon cycle. Uncertainties are large, however (Table 1), and caution remain necessary. Comparative studies of hydrological systems in tropical, temperate and arctic regions presenting a range of sediment discharge and erosion characteristics would be helpful.

Finally, we find that the net budget of the carbon cycle ($\Delta_c$ ~ 3.9 TmolC yr$^{-1}$) is commensurate with the flux of $OC_{petro}$ that escapes reoxidation (ca. 3.6 TmolC yr$^{-1}$). The similarity between $\Delta_c$ and $OC_{petro}$ export suggests that the persistence of an oxygenated surface environment in the Pleistocene may not a problem of *supply* (ie. of reductant and oxidant), but rather one of sluggish global OC oxidation kinetics in the environment. In hindsight, the formation of graphitic materials in rocks, a ubiquitous and complex phenomenon (41, 42), may be the most important yet overlooked regulator of the long-term kinetics of the organic carbon cycle.

/3024 words/

**Figure Captions**

**Figure 1**. Redox exchanges between continents, mantle and ocean-atmosphere system. The redox budget of mantle exchanges with atmosphere is $\Delta^s = \Delta^s_{OC} + \Delta^s_S + \Delta^s_{Fe}$. The budget of continental weathering/marine sediment (deep sea and margin) deposition sub-cycle is $\Delta^c = \Delta^c_{OC} + \Delta^c_S + \Delta^c_{Fe}$. Each $\Delta^c_i = F^{in}_i + F^{out}_i$ represent the budget between continental weathering and continental accretion (including margin sediment flux and accreted deep sediment, cf Table 1 and Supplementary Information). It includes the budget between continent and atmosphere ($F^{c,ac} + F^{c,o}$) and the sediment input to subduction zones ($F^{c,su}$) (cf. supplementary figure 1). Note that this model does not explicitly represent the redox exchange between continents and mantle, which do not directly impact the AO system.

**Figure 2**. **A**. The histogram show the result of the Monte Carlo Simulation of redox output from the Fe cycles (upper half) and S cycles (lower half). We distinguish the sedimentary cycle (right half), and the igneous cycles (left half). Details on the computation can be found in the text, and in Supplementary Material. **B**. Histogram presenting the redox output of the carbon cycle as whole. It is likely to be a net source of oxygen, despite all uncertainties in fluxes. For each panels, an inset presents the main features of the model: element concentration, element fluxes used, ratio between different valence states of an element used for different reservoirs (e.g. $Fe^{3+}/Fe_{tot}$).

**Figure 3**. Redox pathways coupling geological Fe and C cycles. In this illustrative model, the S cycle is not represented for sake of clarity. The net export of about 40 Mt $OC_{bio}$ yr$^{-1}$ from surface to lithosphere is the cost (cf abstract)—(equivalent to $1.5 \times 10^6$ TJ yr$^{-1}$ of photosynthetic (solar) energy (~10000 Nagasaki bombs)— for a near steady-state atmospheric $O_2$ composition to exist far from thermodynamic equilibrium in the Pleistocene.

**Table Captions**

**Table 1.** Global $O_2$ budget (x$10^{12}$ mol $O_2$ equivalent per year, equivalent to 4x$10^{12}$ mol e$^-$ equivalents per year)

$Ᵽ$ represents to burial flux of reduced species (oceanic lithosphere and sediments), $Ᵽ$ represent fluxes removing $O_2$ from the atmosphere-ocean systems (producing of reduced species at MOR and continental weathering fluxes). We converted element flux to equivalent electron and $O_2$ flux by assuming reference valence states of C, Fe, S, and H are +4, +3, +6 and +1 in the atmosphere-ocean system, respectively (Supplementary Information). ¥ standard deviations (σ) are either determined from inventory estimates in the literature, or derived by Monte Carlo analysis # cf Supplementary Material) § cf Supplementary Material. ¶ This flux denotes only $OC_{bio}$ (cf Supplementary Material). The fraction of deep-sea (pelagic) sediments accreted to continents is 0.2 (43) ¤ This flux denotes only $OC_{bio}$. It incorporates $OC_{bio}$ deposited on shelves, pelagic $OC_{bio}$ accreted. The conversion of $OC_{bio}$ burial to $O_2$ equivalent flux assumes a burial quotient $O_2$:$OC_{bio}$ of 1.1 to account for deviation from ideal $CH_2O$ stoichiometry during oxidative diagenesis and reductive thermal maturation (cf Supplementary Material) ¢ from ref (21). Þ represents fluctuation in the size of dissolved, particulate organic carbon and biomass reservoir. £ from ref (1) Ƕ petrogenic OC oxidation corresponds only to oxidation of partly graphitized $OC_{petro}$ in continental rocks. The oxidation of thermogenic methane is a composite flux from (18) and (44), and the methane flux of ~25 Tmol yr$^{-1}$ is consistent with a decarboxylation flux of ~20 Mt yr$^{-1}$ (45)

**Acknowledgements** : This work is supported by a Society in Science, Branco Weiss fellowship grant (MG). Partial support from ETH installation funds to Olivier Bachmann is also gratefully acknowledged.

# Figure 1

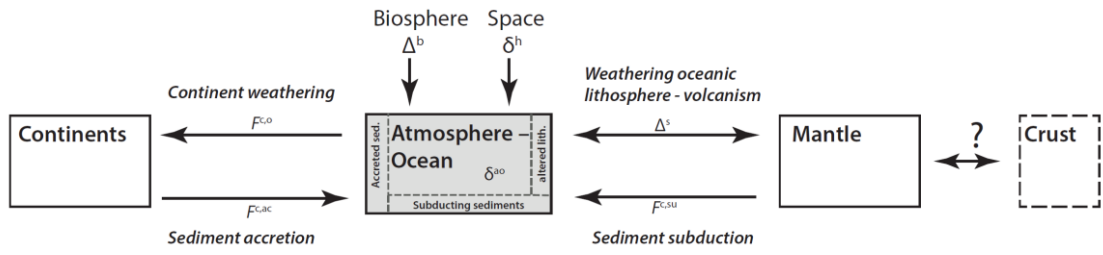

Figure 2

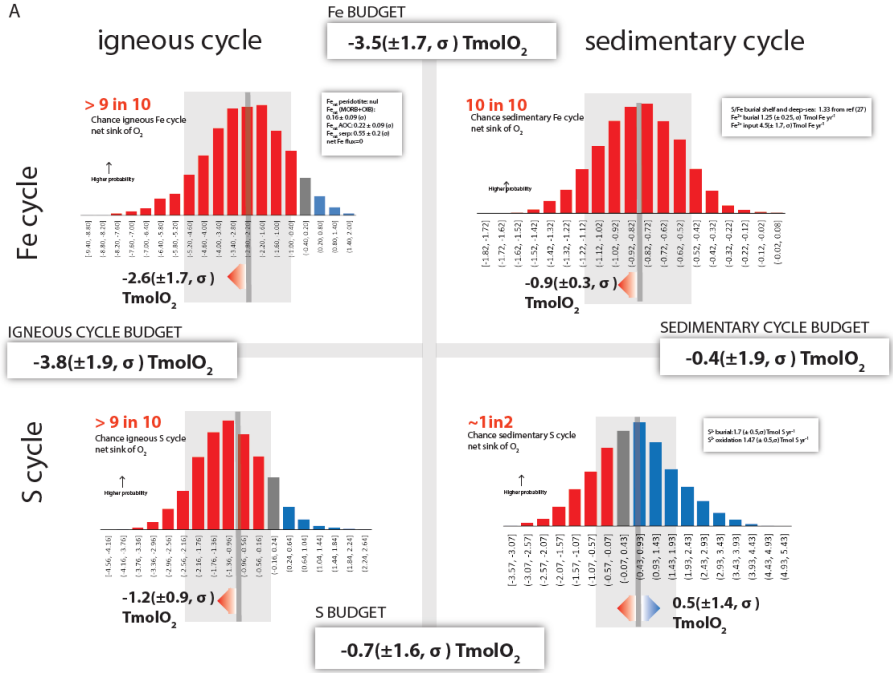

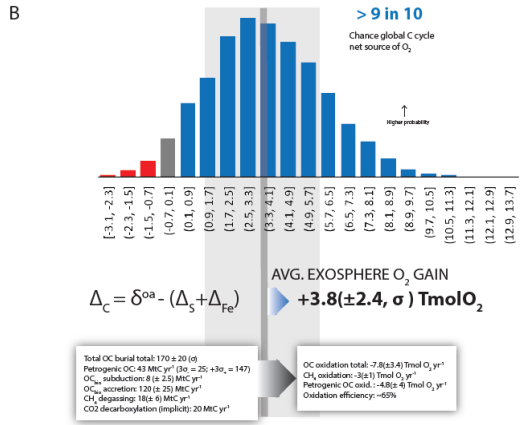

Figure 3

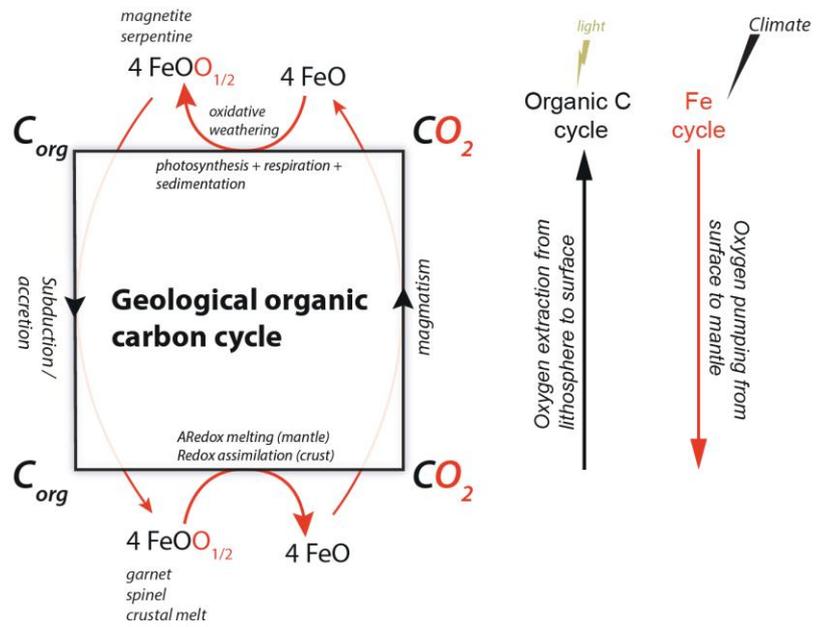

**Table 1. Global O$_2$ budget (x10$^{12}$ mol O$_2$ equivalent per year)**

| Type of Flux | symbol | O$_2$ flux | σ[¥] |
|---|---|---|---|
| **Measured** | | | |
| O$_2$ sinks | | | |
| subduction of oceanic crust and MOR-OIB[#] | $\Delta^s$ | -3.8 | 1.0 |
| continental oxidation of non-C (Fe,S)[§] | $F^{c,o}_{s.fe}$ | -3.2 | 0.5 |
| | | | |
| O$_2$ sources | | | |
| burial in deep sea sediment (Fe,S subduction)[§] | $F^{c,su}_{S.Fe}$ | 0.3 | 0.1 |
| OC$_{bio}$ burial in subducted deep sea sediments (subduction)[¶] | $F^{c,su}_{OC}$ | 0.7 | 0.2 |
| burial in continental margin sediment (Fe,S)[§] | $F^{c,ac}_{S.Fe}$ | 2.5 | 1.4 |
| OC$_{bio}$ burial in accreted sediment[Ħ] | $F^{c,ac}_{OC}$ | 10.8 | 2.4 |
| OC$_{petro}$ burial in accreted sediment[Ħ] | $F_{petro}$ | 3.6 | $3\sigma_- = 2.1$ <br> $3\sigma_+ = 5.1$ |
| H loss in upper atmosphere[c] | $\delta^h$ | ~0.02 | |
| photosynthesis+respiration (POC,DOC,biomass)[b] | $\Delta^b$ | ~0 | |
| AO redox balance[£] | $\delta^{ao}$ | -0.3 | |
| | | | |
| **Derived** | | | |
| continental weathering/deposition balance | $\Delta^c$ | 3.5 | 1.9 |
| Net export by OC cycle | $\Delta^c_{oc}$ | 3.9 | 2.6 |
| total OC oxidation | $F^{c,o}_{OC}$ | -7.6 | 3.5 |
| OC$_{petro}$ oxidation[ʉ] | $F^{c,o}_{OCpetro}$ | -4.6 | 3.7 |
| OC$_{petro}$ oxidation efficiency | $\Gamma_{OCpetro}$ | 64% | |
| | | | |
| Redox budget sulfur cycle (igneous and sedimentary) | $\Delta_s$ | -0.8 | 1.9 |
| Redox budget iron cycle (igneous and sedimentary) | $\Delta_{Fe}$ | -3.4 | 1.8 |
| Redox budget carbon cycle (igneous and sedimentary) | $\Delta_C$ | 3.9 | 2.6 |

Supplementary Material for

# Glaciological Window into the Pace of Earth's Organic Carbon Cycle

Matthieu Emmanuel Galvez

Email: **matthieu.galvez@gmail.com**

**This PDF file includes:**

    Supplementary text

    Tables S1 and its caption

    References for SI citations

*Supplementary Material Text*

1. **Materials and Methods**

    1. Conversion of electron ($e^-$) flux to $O_2$ equivalent ($O_{2,eq}$) flux

We converted element flux to electron flux (moles) by assuming reference valence states of C, Fe, S, and H are +4, +3, +6 and +1 in the atmosphere-ocean system, respectively. The electron flux is the number of moles of electron needed to convert the element in its given valence state, e.g. $S^-$, to its reference valence state, e.g. $S^{6+}$ (ie. 7 moles here).

Electron flux are then converted to molar $O_2$ equivalents by applying the equation

$$O_2 + 4e^- = 2O^{2-} \quad (1)$$

which gives $nO_2 = ne^-/4$

2. **Igneous cycle: mid ocean ridge flux, hydrothermal weathering. And volcanic degassing fluxe**

    2.1. Principle for the calculation of Redox Budgets associated to hydrothermal alteration of the oceanic lithosphere (referred to as the 'igneous cycle')

This procedure computes the net flux of oxidized forms of Fe ($Fe^{3+}$) and S ($S^{6+}$) to the subduction zone. The latter are considered to be the reference valence state of Fe and S in the exogenous reservoirs. We compute it by comparing the oxidation state of the altered oceanic lithosphere (AOL) to that of the background, unaltered lithosphere, ie. the amount of $O_2$ added to the lithosphere during seafloor alteration of igneous Fe and S. For a given amount of iron submitted to alteration, a certain fraction comes back more oxidized than it came in.

    2.2. Total mass flux (valid for computation of Fe, S redox budgets)

The AOL is composed of a top layer of basalt and of a lower layer of gabbros (basalt + gabbro are referred to as altered oceanic crust, AOC) of thickness 2 km and 4 km, respectively(17). The lower section of the AOL is composed of the altered oceanic mantle (AOM, ranging from 0.5 to 3 km, of100% serpentinized mantle). Assuming a lithospheric subduction rate of 2.45 km² yr⁻¹ after ref (26), and densities of 3100 kg m³ for the AOC and of 2800 kg m³ for the AOM, gives a total mass flux of 45.6 × 10¹⁵ g yr⁻¹ for the AOC, and of 13 (±10) x 10¹⁵ g yr⁻¹ for the AOM. Evans (19) recommends a ± 20% uncertainty on the AOC flux. We adopt this value.

### 2.3. Iron: $\Delta^s Fe$

#### 2.3.1. Fe mass flux and $Fe^{3+}/Fe_{tot}$ ratios.

For the top 2 km thick altered basalts, the Fe concentration is 7.35 (± 1.41,σ) wt%, and the Fe ratio is $Fe^{3+}/Fe_{tot}$ = 0.26 (±0.08,σ) from ref (17). For the 4 km thick gabbro layer, the Fe concentration is 5.8 (± 3.6,σ) wt%, with $Fe^{3+}/Fe_{tot}$=0.2 (± 0.08,σ) from (17). Combining the structure of the AOC (cf above) with those values give an average $Fe^{3+}/Fe_{tot}$ in the AOC of 0.22 (± 0.09,σ). The $Fe^{2+}$ and $Fe^{3+}$ molar flux to subduction by the AOC is thus **40.2 (± 9.3,σ) x 10¹² mol yr⁻¹**, and **11.4 (± 5.6,σ) x 10¹² mol yr⁻¹**, respectively.

The concentration of Fe in the hydrothermally altered upper mantle is 6.78 (± 0.34,σ) wt% FeO, after ref (19). Note that this is equivalent to a concentration of 5.3 (± 0.27,σ) wt% atomic Fe. The iron ratio is $Fe^{3+}/Fe_{tot}$ = 0.54 (± 0.2,σ) after ref (19). The resulting $Fe^{2+}$ and $Fe^{3+}$ molar flux in the serpentinite layer is 5.56 (± 4.39,σ) x 10¹² mol yr⁻¹, and 6.70 (±5.24,σ) x 10¹² mol yr⁻¹. Because the uncertainty is large compared to the value itself, we have used a skewed distribution that prevents negative values in our Monte Carlo analysis. Our value for $Fe^{2+}$ is thus **5.56 (3σ₋ = 0; 3σ₊ = 4.39) x 10¹² mol yr⁻¹**, and for $Fe^{3+}$ it is **6.70 (3σ₋ = 0; 3σ₊ = 5.24) x 10¹² mol yr⁻¹**.

The input redox budget, $F^{s,i}_{Fe}$ (Table 1, main text), is the equivalent $O_2$ flux associated to the subduction of $Fe^{2+}$ out of the exosphere. Because AOC and AOM subducts a total of ≈ 45.8 TmolFe²⁺ yr⁻¹, the $O_2$ equivalent is $F^{s,i}_{Fe}$ **= - 45.8 x 10¹² x (-1) / 4 = 11.4 (± 2.5,σ) x 10¹² Tmol yr⁻¹** eq. This input value of our model is reported in Supplementary Table 1.

#### 2.3.2. Redox budget.

The net budget requires estimating the initial composition of the oceanic lithosphere fraction that has been submitted to alteration. It may not be assumed that it was only $Fe^{2+}$ since a fraction of Fe is already oxidized in the MORB and mantle source that it comes.

Because we only focus on exchanges between lithosphere and exosphere, our procedure differs from (19). For consistency of the calculation of electron flux, we use the same total Fe flux than lost to subduction zone, 35.7 x $10^7$ g Fe $yr^{-1}$, to compute the initial $Fe^{2+}$ and $Fe^{3+}$ flux submitted to alteration.

For unaltered MORB, we take $Fe^{3+}/Fe_{tot}$ = 0.16 (± 0.09,σ). This covers the estimate by Birner et al.(46). The $Fe^{3+}$ content of the pristine peridotite is considered negligible. We obtain a background $Fe^{2+}$ and $Fe^{3+}$ input by the unaltered oceanic lithosphere (MORB plus peridotite) of **56.7 x $10^{12}$ Tmol $yr^{-1}$** and **7.2 x $10^{12}$ Tmol $yr^{-1}$** respectively, ie. prior to alteration.

This is equivalent to a sink of $O_2$ $F^{s,o}_{Fe}$ = 56.2 x $10^{12}$ *(-1) / 4 = **14 (± 1.7,σ) x $10^{12}$ Tmol $yr^{-1}$** (Supplementary Table 1). The overall redox budget is therefore $Δ^s_{Fe} = F^{s,i}_{Fe} - F^{s,o}_{Fe}$ = **-2.6 (± 1.7,σ) x $10^{12}$ Tmol $yr^{-1}$** (Supplementary Table 1). The igneous cycle of iron is an oxygen sink, which concurs with previous research(12). The idea here is to confront this knowledge with the glaciological evidence for limited $O_2$ loss over the last 1 Myr.

### 2.4. Sulfur: $Δ^s_S$

Similarly, we estimate the amount of S of the altered lithosphere that subducts in a more oxidized form than it came in. For a given amount of S submitted to alteration, a certain fraction is redeposited as anhydrite during hydrothermal upwheling(47). Our budget assumes that the S flux subducted either comes from the igneous flux (MORB + OIB) after(47).

### 2.4.1. Sulfur mass flux and speciation flux.

For the AOC, we take a total $S^-$ and $S^{6+}$ concentration of 0.0716 wt% (20% uncertainty(19)) and 0.036 wt% (50% uncertainty(19)), respectively(19). The $S^-$ and $S^{6+}$ molar flux in the AOC is thus 1.02 (± 0.28,σ) x $10^{12}$ mol $yr^{-1}$, and 0.5 (±0.28,σ) x $10^{12}$ mol $yr^{-1}$.

In the hydrothermally altered upper mantle we take a total $S^{2-}$, $S^-$ and $S^{6+}$ concentration of 0.041(±0.55,σ) wt%, 0.048(±0.14,σ) wt%, and 0.078(±0.073,σ) wt% respectively(19). The corresponding molar flux of $S^{2-}$, $S^-$ and $S^{6+}$ is 0.17(±0.24,σ) x $10^{12}$ mol $yr^{-1}$, 0.19(± 0.54,σ) x $10^{12}$ mol $yr^{-1}$, and 0.32(±0.38,σ) x $10^{12}$ mol $yr^{-1}$ respectively.

The redox budget of input $F^{s,i}_S$ is the equivalent O₂ flux associated to ouput of $S^{2-}$ and $S^-$ from the exosphere by subduction of oceanic lithosphere, ie. **$F^{s,i}_S$ = - 0.17x10¹² x (-8)/4 - 1.21x10¹²x (-7) / 4 = 2.5 (± 0.5,σ) x 10¹² Tmol yr⁻¹** equivalent O₂.

### 2.4.2. Redox budget.

This output flux is compared to input flux of S by MORB, OIB and arc gases.

MORB content is 0.08± 0.01 wt% as $S^{2-}$ and 0.006 wt% as $S^{6+}$, this corresponds to about 0.09± 0.01 wt% S in MORB(19), with a $S^{6+}/S_{tot}$ ratio of 0.07 after Wallace and Carmichael(19). This corresponds to a $S^{2-}$ and $S^{6+}$ fluxes of 1.2± 0.01 Tmol yr⁻¹ and 0.09 Tmol yr⁻¹, respectively. OIB flux is 4.56 x 10¹⁵ g yr⁻¹, for $S^{2-}$ content is 0.33± 0.1 wt% (0.47 Tmol yr⁻¹) and $S^{6+}$ content of 0.06± 0.02 wt% (0.009 Tmol yr⁻¹). This corresponds to a $S^{2-}$ and $S^{6+}$ fluxes of 0.47± 0.17Tmol yr⁻¹ and 0.08±0.04 Tmol yr⁻¹.

The arc degassing flux is 0.32± 0.06 Tmol S yr⁻¹ (as SO₂) after Hilton Fischer and Marty (48). Thus, the total mantle input flux of S to exosphere (in all its valence forms) is ~2.15 TmolS yr⁻¹, which balances the subduction zone output of ~2.21 TmolS yr⁻¹ within the margin of error. This supports the view that hydrothermal systems are operating as near-close systems as assumed by Tostevin(29). The difference of about 0.1 Tmol suggest minimal uptake from seawater, which is consistent with Alt 2013(49).

The total $S^{2-}$ input flux is 1.66±0.29 Tmol yr⁻¹, and the total $S^{4+}$ input flux is 0.32±0.06 Tmol yr⁻¹. Overall, the input flux of reduced S is a sink of O₂ of **$F^{s,o}_S$ = 1.66x10¹² x (-8)/4 + 0.32x10¹² x (-2)/4 = -3.5 TmolO₂ yr⁻¹** (Supplementary Table 1).

### 2.5. Hydrogen: Δ$^s$H

We use avolcanic H⁰ flux of 0.17 Tmol yr⁻¹, corresponding to a ratio $H^{1+}/H^0$ = 1/10 from(12), with $H^{1+}$ flux ((H₂O) of about 17 Tmol yr⁻¹. We adopt this value which corresponds to an O₂ sink of 0.17 Tmol*(-1)/4 = **$F^{s}_H$ = -0.04 TmolO₂ yr⁻¹**.

## 3. Continental cycle: marine sediments (margins and deep sea) and continental weathering fluxes.

### 3.1. Principle

All marine sediments, subducted or accreted, are part of the continental cycles of Fe, S and C. This is an arbitrary decision, based on the idea that those sediments are intrinsically continental materials, wherever they end up in accretionary prisms. Our budget is independent of this choice. The continental cycle needs to be a net source of oxygen to balance the oceanic sinks (Main text).

### 3.2. Carbon: $\Delta^c c$

#### 3.2.1. C burial flux: $F^{c,i}_{OC}$

##### 3.2.1.1. Total OC burial flux

The total OC burial flux is composed of biological material ($OC_{bio}$) and petrogenic material ($OC_{petro}$) of marine and terrestrial sources(33, 50, 51). In his original work, Berner(45) estimated a global C burial flux of about 121 MtC yr$^{-1}$. But this flux assumes a diagenetic loss of about 20% of the initial C deposited on the sea-floor (about 30 MtC yr$^{-1}$) by decarboxylation, ie. loss of $CO_2$ (cf ref(45)). Later, the updates by Hedges and Keil(52) and Smith et al.(34) have been applied to this 'decarboxylated' flux and not to the primary OC deposition flux of Berner(45) of 162 Mt. We consider, by security, that the latest estimates of about 170 Mt/yr from ref(34) implicitly assumes about 20% initial loss due to decarboxylation. Thus, we attribute a large error to this flux(53): we take here a total OC burial flux of 170(± 20, σ) Mt C yr$^{-1}$, with 160(± 20, σ) Mt C yr$^{-1}$ buried in margin sediments, and 10(± 3, σ) Mt C yr$^{-10}$ buried in deep sea 'pelagic' sediments (35). Clift has shown that about 20% of the pelagic flux ends up accreted to continents over the last 10 Myr, whereas 80% is accreted(54). We use this value to segregate the total burial flux between subduction to mantle and accretion to continents.

##### 3.2.1.2. Estimated $OC_{petro}$ and $OC_{bio}$

This total OC flux contains a fraction derived from eroded rocks on continents, $OC_{petro}$, which does not contribute to the redox budget. The reburied $OC_{petro}$ fraction is $F_{petro}$ = 3.6 (3σ$_-$ = 2.1; 3σ$_+$ = 5.1) TmolC yr$^{-1}$ [from ref (33)]. This amount is subtracted from the total

OC burial flux to obtain the total $OC_{bio}$ fraction subducted [0.67(± 0.22, σ) TmolC yr$^{-1}$] and accreted [9.84(± 2.21, σ) TmolC yr$^{-1}$].

Another immediate consequence of the uncertainty of decarboxylation processes described above is that the burial of decarboxylated kerogens, ie. 0.67 + 9.84 = 10.53(± 2.43, σ) TmolOC$_{bio}$ yr$^{-1}$, releases more $O_2$ per mole of buried C than the $CH_2O$ molecule would. Therefore, we assume that each mole of OCbio buried releases a little bit more O2 than CH2O would: the burial quotient, adapted from refs (55), $O_2$:C is 1.1. Thus, this procedure accounts for deviation from ideal $CH_2O$ stoichiometry during oxidative diagenesis and reductive thermal maturation, to produce elemental C. The loss corresponds to ~20 MtC yr$^{-1}$ lost to the ocean by decarboxylation during diagenesis. We note that this flux is equivalent to the flux of methane lost during catagenesis [here 1.5 (±0.5, σ) Tmol $CH_4$ yr$^{-1}$, see below 3.2.2].

The $O_2$ equivalent for the global $OC_{bio}$ burial ~ 10.54 Tmol yr$^{-1}$ is obtained by mutiplying it by 1.1, yielding **$F^{c,i}_{OC}$ = -11.6 (±2.5, σ) Tmol $O_2$ yr$^{-1}$**.

3.2.2. $OC_{petro}$ and $CH_4$ oxidation flux

The total OC oxidation flux $F^{c,o}_{OC}$ is derived from our redox budget above, as described in the main text. We obtain $F^{c,o}_{OC}$ = -7.7 (±3.5, σ) Tmol $O_2$ yr$^{-1}$. $F^{c,o}_{OC}$ is split between thermogenic $CH_4$ oxidation ($F^{c,o}_{ch4}$) and $OC_{petro}$ oxidation ($F^{c,o}_{petro}$). If $F^{c,o}_{ch4}$ = -3 (±1, σ) Tmol $O_2$ yr$^{-1}$ (Table 1, ref (18)] to cover the broad range of independent estimates [cf. ref(18)], we obtain by difference $F^{c,o}_{petro}$ = -4.8 (±3.6, σ) Tmol $O_2$ yr$^{-1}$ (Table 1).

3.3. Sulfur: $\Delta^c S$

3.3.1. Sulfide weathering

Laakso considered 20% of sulfate flux to ocean is derived from sulfide weathering, a canonical assumption since Berner and Lerman and Garrels(56, 57). Recent riverine S stable isotopes of 4.8‰ suggest otherwise(10). The authors suggest that the fraction of weathered sulfide in the riverine sulfate is about 42%(10). Considering a total weathered S flux of 2.8 (±0.5, σ)

Tmol S yr$^{-1}$ gives a sulfide (pyrite) weathering flux of 1.3 (±0.5, σ) Tmol S yr$^{-1}$, considered to be S in its -1 valence state (S$^-$) for conversion to O$_2$ equivalents.

The O$_2$ equivalent is $F^{c,o}_S$ = **1.3 *(-7) / 4 = -2.06 (±2.5, σ) Tmol O$_2$** yr$^{-1}$ (Supplementary Table 1).

### 3.3.2. Sulfide burial

Another recent isotopic study(29) suggests that about 20-35% of the total flux should precipitate as sulfide in marine sediments, ie. about 0.6 to 1.1 Tmol S yr$^{-1}$. This flux is on the low end of the total subducted and accreted S flux used here of **1.4 Tmol O$_2$** yr$^{-1}$. The latter is obtained from our OC$_{bio}$ burial flux (cf. 3.2.1.2) of 0.67 + 9.84 = 10.53(± 2.43, σ) TmolC yr$^{-1}$ using a molar C/S ratio of 7.48 from Berner (45). Our flux is, however, consistent with the S burial flux estimated by Canfield of ca 1.7 Tmol S yr$^{-1}$(27). To cover this large range of uncertainty, we take σ = 0.8 Tmol S yr$^{-1}$ on our S$^-$ burial flux.

The O$_2$ equivalent is $F^{c,i}_S$ = **-1.4 *(-7) / 4 = 2.43(±1.4, σ) Tmol O$_2$** yr$^{-1}$ (Supplementary Table 1).

### 3.3.3. Redox budget

We obtain

$\Delta^c S$ = $F^{c,o}_S$ + $F^{c,i}_{OC}$ = **0.45 (±1.7, σ) Tmol O$_2$ yr$^{-1}$** (Supplementary Table 1).

### 3.4. Iron: $\Delta^c Fe$

Fe weathering and burial.

There is even less constraints on the Fe cycle. Here we take a S:Fe molar ration of 2, which assumes Fe burial is controlled by pyrite precipitation. This gives an Fe burial flux of 1.2(±0.4, σ) TmolFe$^{2+}$ yr$^{-1}$, and $F^{c,i}_{Fe}$ = **- 1.2 * (-1) /4 = -0.3 (± 0.07, σ) Tmol O$_2$ yr$^{-1}$** (Supplementary Table 1).

Estimates of the Fe weathering flux are unclear, and we take her a flux of $F^{c,i}_{Fe}$ = **4.5 * (-1) /4 = -1.12 (± 0.42, σ) Tmol O$_2$ yr$^{-1}$** from (28) (Supplementary Table 1).

Overall, we find that the sedimentary cycles of Fe and S are collectively close to redox neutrality, with $\Delta^c_{S,Fe}$= -0.4 (± 1.4, σ) Tmol $O_2$ $yr^{-1}$, which is consistent with ref (16).

Supplementary Table 1

| | | | | value | | σ |
|---|---|---|---|---|---|---|
| $\Delta^t$ | | | | -3.8 | ± | 1.9 |
| | $\Delta^t_S$ | | | -1.2 | ± | 0.9 |
| | $\Delta^t_{Fe}$ | | | -2.6 | ± | 1.7 |
| | $\Delta^t_C$ | | | | ± | 0.0 |
| $\Delta^c$ | | | | 3.5 | ± | 1.9 |
| | $\Delta^c_S$ | | | 0.4 | ± | 1.7 |
| | $\Delta^c_{Fe}$ | | | -0.8 | ± | 0.4 |
| | $\Delta^c_C$ | | | 3.9 | ± | 2.6 |
| $\Delta_S$ | | | | -0.8 | ± | 1.9 |
| $\Delta_{Fe}$ | | | | -3.4 | ± | 1.8 |
| $\Delta_C$ | | | | 3.9 | ± | 2.6 |
| | | | | | | |
| $\Delta^t$ | | | | -3.8 | ± | 1.9 |
| $\Delta^t_S$ | | | | -1.2 | ± | 0.9 |
| | $F^{t,i}_S$ | | | 2.4 | ± | 0.7 |
| | $F^{t,o}_S$ | | | -3.6 | ± | 0.6 |
| $\Delta^t_{Fe}$ | | | | -2.6 | ± | 1.7 |
| | $F^{t,i}_{Fe}$ | | | 11.4 | ± | 2.5 |
| | $F^{t,o}_{Fe}$ | | | -14.1 | ± | 3.0 |
| $\Delta^t_C$ | | | | -0.1 | ± | 0.0 |
| | $F^{t,i}_{OC}$ | | | -0.1 | ± | 0.0 |
| | $F^{t,o}_{OC}$ | | | 0.0 | ± | 0.0 |
| $\Delta^c$ | | | | 3.5 | | 1.9 |
| $\Delta^c_S$ | | | | 0.4 | ± | 1.7 |
| | $F^{c,i}_S$ | | | 2.5 | ± | 1.4 |
| | | $F^{c,ac}_S$ | | 2.3 | ± | 1.4 |
| | | $F^{c,su}_S$ | | 0.2 | ± | 0.0 |
| | $F^{c,o}_S$ | | | -2.1 | ± | 0.9 |
| $\Delta^c_{Fe}$ | | | | -0.8 | ± | 0.4 |
| | $F^{c,i}_{Fe}$ | | | 0.3 | ± | 0.1 |
| | | $F^{c,ac}_{Fe}$ | | 0.2 | ± | 0.1 |
| | | $F^{c,su}_{Fe}$ | | 0.1 | ± | 0.0 |
| | $F^{c,o}_{Fe}$ | | | -1.1 | ± | 0.4 |
| $\Delta^c_C$ | | | | 3.9 | ± | 2.6 |
| | $F^{c,i}_{OC}$ | | | -11.6 | ± | 2.4 |
| | | $F^{c,ac}_{OC}$ | | 10.8 | ± | 2.4 |
| | | $F^{c,su}_{OC}$ | | 0.7 | ± | 0.2 |
| | $F^{t,o}_{OC}$ | | | -7.7 | ± | 3.6 |
| | | $F^{c,o}_{ch4}$ | | -3.0 | ± | 1.0 |
| | | $F^{t,o}_{OCpetro}$ | | -4.7 | ± | 3.7 |